# Towards a transportable Ca$^+$ optical clock with a systematic uncertainty of 4.8×10$^{-18}$

Mengyan Zeng,[3,1,2] Yao Huang,[1,2] Baolin Zhang,[1,2] Yanmei Hao,[1,2,4] Zixiao Ma,[1,2,4] Ruming Hu,[1,2,4] Huaqing Zhang,[1,2] Zheng Chen,[1,2,4] Miao Wang,[1,2] Hua Guan,[1,2,5] [*] Kelin Gao,[1,2] [†]

[1]*State Key Laboratory of Magnetic Resonance and Atomic and Molecular Physics, Innovation Academy for Precision Measurement Science and Technology, Chinese Academy of Sciences, Wuhan 430071, China*
[2]*Key Laboratory of Atomic Frequency Standards, Innovation Academy for Precision Measurement Science and Technology, Chinese Academy of Sciences, Wuhan 430071, China*
[3]*Huazhong University of Science and Technology, Wuhan 430074, China*
[4]*University of Chinese Academy of Sciences, Beijing 100049, China*
[5]*Wuhan Institute of Quantum Technology, Wuhan 430206, China*

We present a compact, long-term nearly continuous operation of a room-temperature Ca$^+$ optical clock setup towards a transportable clock, achieving an overall systematic uncertainty of $4.8 \times 10^{-18}$ and an uptime rate of 97.8% over an 8-day period. The active liquid-cooling scheme is adopted, combined with the precise temperature measurement with 13 temperature sensors both inside and outside the vacuum chamber to ensure the accurate evaluation of the thermal environment for the optical clock. The environmental temperature uncertainty is evaluated as 293.31(0.4) K, corresponding to a blackbody radiation (BBR) frequency shift uncertainty of $4.6 \times 10^{-18}$, which is reduced more than two times compared to our previous work. Through the frequency comparison between the room temperature Ca$^+$ optical clock and a cryogenic Ca$^+$ optical clock, the overall uncertainty of the clock comparison is $7.5 \times 10^{-18}$, including a statistic uncertainty of $4.9 \times 10^{-18}$ and a systematic uncertainty of $5.7 \times 10^{-18}$. This work provides a set of feasible implementations for high-precision transportable ion optical clocks.

## I. INTRODUCTION

Optical clocks based on neutral atoms trapped with optical lattice or single ions trapped with ion traps have been greatly improved in the past two decades [1-7]. At present, the systematic uncertainty of optical clocks referenced to Yb [3], Sr [2,4], Al$^+$ [5], Yb$^+$ [6] and Ca$^+$ [7] have reached to the $10^{-18}$ level or below [5]. The uncertainties of the above state-of-the-art optical clocks have all exceeded the best Cs fountain clocks by more than one order of magnitude [8], with promising applications ranging from the redefinition of the SI second [8,9], testing of the general relativity [10,11], testing if fundamental constants vary through time and space [11-14], the geoid measurements [15], etc. The BIPM has proposed a roadmap for the redefinition of the SI second in 2016, which clearly stated that it is necessary to develop the transportable optical clock with an uncertainty at the 10$^{-18}$ level, in order to measure the frequency ratios between optical clocks in different institutes, with uncertainties at the 10$^{-18}$ level, for checking the reproducibility of the optical clocks and their frequency comparisons [8]. Besides, in terms of geoid measurements, a transportable optical clock with an uncertainty at the 10$^{-18}$ level is required to achieve a centimeter-level height difference resolution [16]. This will surpass the resolution for traditional geodetic measurement scheme, which would demonstrate the advantages of the optical-clock-based geopotential measurement. However, it is challenging to realize a transportable, high-performance optical clock. Lots of institutions all over the world are now carrying out researches on transportable optical clocks [16-20]. Among them, transportable Sr and Ca[+] optical clocks assembled in the car trailers with systematic uncertainties at the 10[-17] level groups have been developed [17,19]. RIKEN of Japan has developed a cryogenic transportable Sr optical lattice clock at 245 K to suppress the BBR shift uncertainty, achieving a systematic uncertainty of $5.5 \times 10^{-18}$ [16].

The Ca[+] ion is an ideal reference candidate for building a transportable optical clock. It has a relatively simple level scheme, requiring fewer numbers of diode lasers with weaker power compared to the Al[+]-, Yb[+]- based optical clocks [5,6] and the Yb-, Sr- based optical lattice clocks [3,4], allowing for building a compact, low-cost and reliable transportable optical clock [19,20]. The uncertainty of our previous transportable Ca[+] ion optical clocks is $1.3 \times 10^{-17}$, which limited by the BBR shift uncertainty [19]. A liquid nitrogen-cooled cryogenic Ca[+] ion optical clock has been built with a systematic uncertainty of $3.0 \times 10^{-18}$ [7], but it's difficult to be applied to portable optical clocks at the moment. Therefore, reducing the uncertainty of BBR shift at room temperature is an important goal for building 10[-18] level uncertainty transportable Ca[+] optical clocks. In this work, temperature stabilization of the thermal environment, precise temperature measurement and evaluation schemes are introduced to reduce the temperature variation and improve the accuracy of temperature evaluation. The uncertainty for the BBR frequency shift evaluation has then been lowered to the 10[-18] level, and a frequency comparison with an overall uncertainty of $7.5 \times 10^{-18}$ has been implemented for the verification of the uncertainty evaluation.

In addition to the uncertainty, the continuous operation ability is the other essential feature of an optical clock for redefining the second [8]. In this paper, we report on an over-one-week operation of a room-temperature Ca[+] optical clock with an improved uptime rate of 95%, after the optimization of the laser system and the control program. Compared to other ion-based optical clocks, a group have reported the long-term quasi-continuous performance of an Yb[+] clock with 76% uptime for 25 days [29]. Regarding optical lattice clocks, Sr optical lattice clocks with uptime rates of 93% for 10 days [30] and 84% for 25 days [31] have been reported, a long-term operation of an [171]Yb optical lattice clock for half a year has been reported, including uptimes of 93.9% for the first 24 days [32].

## II. EXPERIMENTAL APPARATUS

Here, we present a room-temperature Ca[+] optical clock with an optimized structure of the ion trap and miniaturized titanium vacuum chamber. A diamond-wafer-based linear ion trap is introduced with low heating rates. The ion trap is built of a laser-machined 300 μm thick diamond wafer, which is gold-plated to generate the radiofrequency (RF) trap electrodes and DC compensation electrodes on the diamond surface. The distances between the trapped ion and RF electrodes are ~0.25 mm. The ion trap wafer and polished titanium endcaps are mounted to the silver-plated oxygen-free copper support with high thermal conductivity, which is attached to an oxygen-free copper heat sink with most surfaces exposed to the external vacuum environment. The heating rates of the trap has been measured as less than 5 quanta/s in all directions [33]. With a probe time of ~ 80 ms, the thermal motion shift uncertainty due to the heating rate is as low as the 10[-19] level.

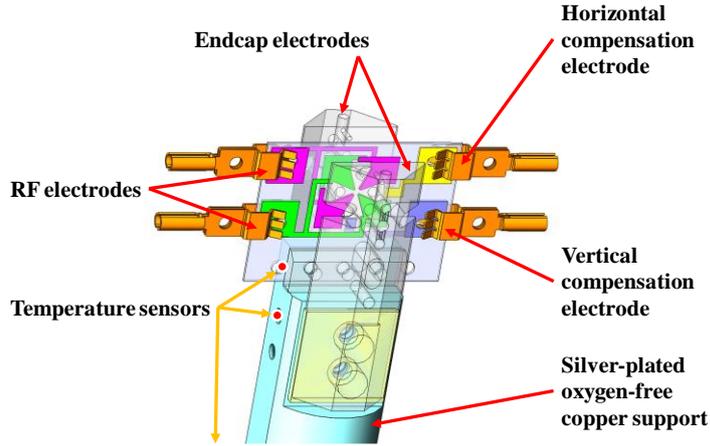

Fig. 1. Simplified schematic of the ion trap system. The design is based on the one used in NIST Al$^+$ clocks [5], with some minor modifications. Three in-vacuum thermocouple sensors with the calibration accuracy of 0.1 K are installed inside the vacuum chamber to characterize the thermal environment; one is installed at the diamond wafer to monitor its fluctuation, and two others are mounted on the top and middle of the silver-plated oxygen-free copper support to measure the temperature fluctuation and gradient of the vacuum chamber. The third thermocouple located in the middle of the support is not shown in the figure. The insulation materials between the endcap electrodes and the copper support are also diamonds. The design makes them nearly invisible to the ions, which means that the BBR temperature uncertainty contributed from insulation supports is greatly suppressed.

The vacuum chamber has a volume of $0.28 \times 0.16 \times 0.19$ m$^3$ and is made of titanium alloy. To achieve a stable magnetic field environment for Ca$^+$, four layers of magnetic shields are added. The vacuum chamber and the optics are placed on a $0.9 \times 0.7$ m$^2$ size table. The table is surrounded by a liquid-cooled box to isolate from the room temperature fluctuations. The size of the box is $1.14 \times 0.84 \times 1.03$ m$^3$. 10 platinum resistance temperature detectors (PRTDs) are installed to evaluate the thermal environment outside the vacuum chamber. The overall temperature measurement system was calibrated with an accuracy level better than 0.1 K at the Hubei Institute of Measurement and Testing Technology. A detailed description of the liquid-cooled box is presented in the section 1 of the Supplemental Material.

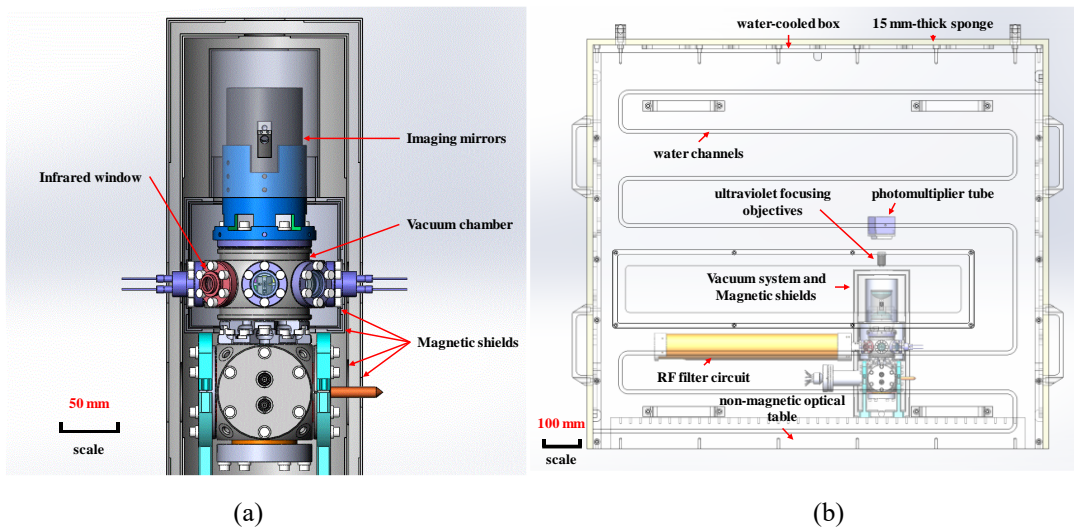

Fig. 2. (a) The vacuum chamber is made of titanium alloy to avoid the magnetic field produced by the chamber. The titanium alloy material also helps to achieve higher vacuum. One CF35 window is on the top for image collection and vertical laser probing, six CF16 windows are on the sides for laser cooling/probing in the corresponding directions, including an infrared window (colored in red) for the temperature measurement with an infrared camera, two CF16 feedthroughs are used for applying voltages on the trapping rf and compensation electrodes. Four layers of magnetic shields are added for significantly weaken of the environmental magnetic field fluctuations. Within the magnetic shields, three pairs of coils are installed for adjusting the strength and direction of the magnetic field. (b) The vacuum chamber is located in a liquid-cooled box to ensure the uniformity and stability of thermal environment. The brown tube in the figure is the helical resonator used for RF supply of the ion trap system. Modules in the system are connected by optical fibers and cables with the laser-stabilization system and circuit control system in the laboratory environment.

Compared to the room-temperature optical clock, the vacuum system of our liquid-nitrogen cryogenic optical clock is relatively complex and huge [7]. Such a complex vacuum system is prone to vacuum leakage during transportation. Moreover, current liquid-nitrogen $Ca^+$ optical clock requires daily liquid nitrogen replenishment to maintain the operation, which prevents it from long-distance transport. In contrast to the cryogenic clock, the vacuum system of the room-temperature optical clock is relatively small and robust for easier transportation. Meanwhile, in the room-temperature optical clock, the liquid-cooled box is simply composed of five liquid-cooled panels, which can be easily disassembled and reassembled. In our preliminary testing for transportation, the clock has been moved from the 1-m-high optical table to the ground, it takes less than a day for the recovery of the clock operation. Therefore, the room-temperature optical clock with our liquid-cooled scheme can be conveniently mounted into our transportable car trailer in the future.

## III. FREQUENCY-SHIFT EVALUATION

Similar to our previously built optical clocks [7,19,23,33], cooling lasers under optimized operation conditions are adopted for reducing the $2^{nd}$-order Doppler shift due to the secular motion of the ion. Considering the trap heating rates, the ion temperature is estimated as 1.19(33) mK during the clock comparison experiments, corresponding to a $2^{nd}$-order Doppler shift (due to secular motion) of $-(4.1 \pm 1.2) \times 10^{-18}$. The excess-micromotion-induced second-order Doppler shift and Stark shift, the combination of the Stark shift due to the secular motion and the second-order Doppler shift due to the secular motion (micromotion induced) have been canceled by choosing the magic rf trapping frequency [23,35]. The excess-micromotion-induced shift and its uncertainty are evaluated as less than $1 \times 10^{-19}$. The quadrupole shift, the first-order Zeeman shift, and the tensor Stark shifts due to the ion motion and the lasers have been canceled by averaging the three pairs of Zeeman components [1,33]. A specific magnetic field direction is chosen to achieve a magic angle between the electric field gradient and the magnetic field direction for the reduction of the electric quadrupole shift drift [7], which greatly suppressed the residual quadrupole shift, with an evaluated uncertainty of $4 \times 10^{-19}$. The servo shift and uncertainty are inferred from the small residual errors in the quantum jump imbalances recoded by the software [33]. The upper limit of the servo-induced frequency shift uncertainty is evaluated to be $4 \times 10^{-19}$. In our experiment, two laser beams in opposite directions are used for independent and interleaved probing of the ion [5,7]. By averaging the two independent frequency measurements, the $1^{st}$-order Doppler shift have been reduced with an uncertainty of $3 \times 10^{-19}$. The ac Stark shift due to the probe laser has been greatly reduced with an uncertainty of < 1×10⁻¹⁹ by adopting the hyper-Ramsey interrogation scheme [34]. More detailed methods and results of the evaluation of the systematic shifts and uncertainties are shown in the section 2 of the Supplemental Material. After the above-mentioned improvements and evaluations, for Ca⁺ optical clocks, the total systematic uncertainty is mainly limited by the BBR shift.

For optical clocks, the BBR shift can be expressed as [21]:

$$\Delta \nu_{BBR} = -\frac{\Delta \alpha_0}{2h} \langle E^2 \rangle_T [1 + \eta(T)] \qquad (1)$$

where $\Delta \alpha_0$ is the differential static scalar polarizability for the clock transition, $\eta(T)$ is the dynamic correction coefficients, $h$ is the Plank's constant, $\langle E^2 \rangle_T = [8.319430(15)V/cm]^2(T/300K)^4$ is the mean-squared electric field in a BBR environment at the temperature $T$ [22]. In our previous work, the differential static scalar polarizability $\Delta \alpha_0$ has been measured as $-7.2677(21) \times 10^{-40} Jm^2V^{-2}$ and the uncertainty of BBR shift contributed by the BBR coefficient $\Delta \alpha_0$ including dynamic correction is reduced to $3 \times 10^{-19}$ [23]. Therefore, the key to reducing the BBR frequency shift uncertainty of the Ca⁺ optical clock at room temperature is the suppression of BBR temperature uncertainty.

In order to estimate the blackbody radiation temperature as accurately as possible, active temperature stabilization by liquid cooling, the precisely measurement of the thermal environment, and the finite element (FE) analysis of the thermal radiation of the setups are implemented in this work. As the Ca⁺ is illuminated by the thermal radiation of the components of the ion trap system, the effective BBR temperature $T_{eff}$ felt by the ion can be calculated in terms of the effective solid angles and the temperatures of the surfaces of all components [24]:

$$T_{eff}^4 = \frac{c}{4\sigma} u = \sum_i \left(\frac{\Omega_i^{eff}}{4\pi}\right) T_i^4 \qquad (2)$$

where $c$ is the speed of light, $u$ is the local field energy density, and $\sigma$ is the Stefan–Boltzmann constant. The radiating surface of each component of the trap system can be described by its own temperature $T_i$ and the effective solid angle $\Omega_i^{eff}$ observed by the ions, and $i$ runs over all enclosure surfaces. The $\Omega_i^{eff}$ represents the weight of the corresponding part surface $i$ when calculating $T_i$.

Analysis of the BBR temperature based on the finite element method shows that large effective solid angle of insulation supports, as well as temperature fluctuation and gradient of vacuum chamber typically give the largest contributions to the BBR temperature uncertainty [25,26]. The diamond-wafer-based linear ion trap configuration makes the effective solid angle contribution of the insulation supports almost disappear. Through active temperature control of the operating thermal environment of the optical clock by liquid cooling, the temperature fluctuation of the vacuum system is reduced from ±1.6 K to ±0.3 K.

We calculated the effective solid angle of each component in the ion trap system by finite element (FE) analysis using the method similar to the one given in Ref. [7] with the COMSOL Multiphysics software. In the model, the calcium ion was replaced by a small blackbody sphere, which acts as a probe. Considering the temperature control of the liquid-cooled box, the absolute temperatures of our vacuum parts such as the vacuum chambers and ion trap are very close to the room temperature, effective solid angles $\Omega_i^{eff}$ in the case of uniform temperature distribution can be used to characterize the realistic

system [27]:

$$\frac{\Omega_i^{eff}}{4\pi} = \left(\frac{T_{eff}}{T_i}\right)^3 \frac{\partial T_{eff}}{\partial T_i} \qquad (3)$$

where $T_{eff}$ is the temperature of the probe, $T_i$ is the set temperature for component *i*. In our case, $T_{eff}/T_i \approx 1$. Typical emissivity values were adopted for different surface materials in the calculations [25], the emissivity values of the endcaps, walls of chamber, ion trap and fused-silica windows are set to be 0.166, 0.166, 0.05 and 0.75, respectively. For comparison, the geometric solid angles have also been calculated using emissivity values of 1 for all surfaces. The results are listed in Table I. The temperature of different components can be measured by the temperature sensors combined with the infrared camera. The temperature measurement details of each component are listed in table II.

Table I. Percentages of the effective and geometric solid angles for different parts of the ion trap system.

| Vacuum parts | Material | Emissivity | $\Omega_i^{eff}/(4\pi)$ | $\Omega_i^{geo}/(4\pi)$ |
|---|---|---|---|---|
| Endcaps | Titanium | 0.166 | 47% | 51% |
| Inner walls of vacuum chamber | Titanium | 0.166 | 26% | 28% |
| Ion trap (including the compensation electrodes) | Gold | 0.05 | 14% | 16% |
| Glass window | Fused silica glass | 0.75 | 13% | 5% |

Table II. Temperature evaluation results for the ion trap system.

| Components | Temperature evaluation value (K) | Uncertainty due to sensor calibration (K) | Uncertainty due to temperature gradient (K) | Uncertainty due to ambient temperature fluctuations (K) | Total Uncertainty (K) |
|---|---|---|---|---|---|
| Endcaps | 293.35 | 0.1 | 0.1 | 0.3 | 0.4 |
| Inner walls of vacuum chamber | 293.25 | 0.1 | 0.2 | 0.3 | 0.4 |
| Ion trap (including the compensation electrodes) | 293.35 | 0.1 | 0.1 | 0.3 | 0.4 |
| Glass window | 293.25 | 0.1 | 0.2 | 0.3 | 0.4 |
| $T_{eff}$ | 293.31 | | | | 0.4 |

The effective BBR temperature felt by the ion is calculated to be 293.31(40) K, corresponding to a BBR shift uncertainty of 1.9 mHz. Finally, the temperature-associated BBR shift uncertainty is determined to be $4.6 \times 10^{-18}$. It can be furtherly reduced by calibrating the temperature sensor with an accuracy of less than 15 mK and designing a more uniform and stable temperature control system for the optical clocks.

Table III summarizes the systematic uncertainty budget for the room temperature clock (clock1

columns). The total systematic uncertainty is $4.8 \times 10^{-18}$, limited by the BBR temperature evaluation precision.

Table III. Systematic shifts and uncertainties for the evaluations of the two clocks and their comparison. Here only the effects with systematic shifts or uncertainty $>1\times10^{-19}$ are shown.

| Contributor ($10^{-18}$) | Fractional systematic shift of clock1($10^{-18}$) | Fractional systematic uncertainty of clock1($10^{-18}$) | Fractional systematic shift of clock2($10^{-18}$) | Fractional systematic uncertainty of clock2($10^{-18}$) | Fractional systematic shift of clock1-clock2($10^{-18}$) | Fractional systematic uncertainty of clock1-clock2($10^{-18}$) |
|---|---|---|---|---|---|---|
| BBR field evaluation (temperature) | 842.8 | 4.6 | 7.3 | 2.7 | 835.5 | 5.3 |
| BBR coefficient ($\Delta\alpha_0$) | 0 | 0.3 | 0 | 0.3 | 0 | 0.4 |
| Excess micromotion | 0 | 0 | 0 | 0.2 | 0 | 0.2 |
| 2$^{nd}$-order Doppler (thermal) | -4.1 | 1.2 | -3.1 | 0.9 | -1 | 1.5 |
| Residual quadrupole | 0 | 0.4 | 0 | 0.4 | 0 | 0.6 |
| Servo | 0 | 0.4 | 0 | 0.4 | 0 | 0.6 |
| 1$^{st}$-order Doppler | 0 | 0.3 | 0 | 0.3 | 0 | 0.4 |
| Gravitational shift | | | | | -27.5 | 1 |
| Total | 838.7 | 4.8 | 4.2 | 3.0 | 807.0 | 5.7 |

## IV. CLOCK COMPARISON

We further measured the frequency difference between the room-temperature Ca$^+$ ion optical clock (clock1 in Table III) and the cryogenic clock (clock2 in Table III, with an uncertainty of $3.0 \times 10^{-18}$) [7] to verify the performance of the room-temperature clock. The two clocks have a height difference of 25(1) cm, corresponding to a frequency shift of 12.4(4) mHz or $27.5(1.0) \times 10^{-18}$. In the frequency comparison experiment, the same clock laser synchronously probed the two clocks to observe the hyper-Ramsey spectroscopies with a free evolution time of 40 ms, and the frequency difference is obtained by comparing the two AOMs used to modulate the two clock laser's frequencies to make them be resonant to their clock transitions. As shown in Table III, a fractional frequency shift difference of 8.070(57) $\times 10^{-16}$ is evaluated, corresponding to a 331.7(2.3) mHz difference and a systematic uncertainty of $5.7 \times 10^{-18}$. As shown in Fig.3, the measured frequency difference is 329.9(2.0) mHz, which is in good agreement with the evaluated value. More details about the clock comparison including the Allan deviation are shown in the section 3 of the Supplemental Material.

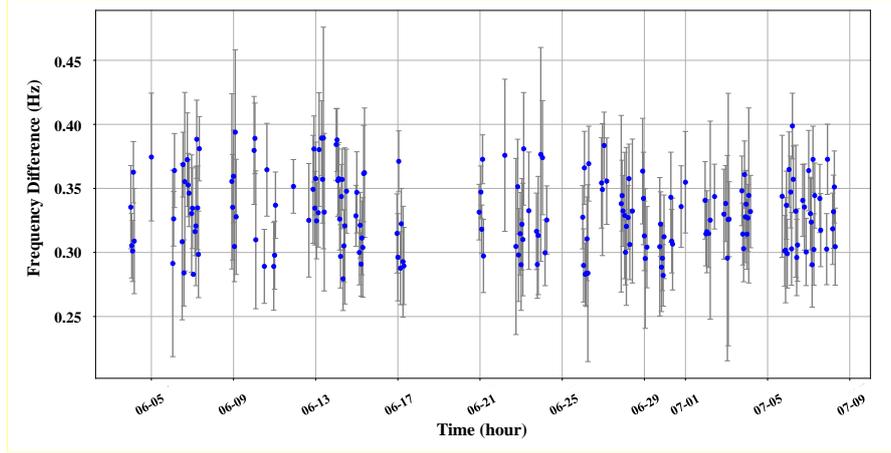

Fig. 3. Frequency comparison of the two clocks. The micromotion minimization process is carried out and the ion temperature for the room-temperature clock is measured daily during the clock comparison. The frequency difference is measured as 329.9 mHz with a statistical uncertainty of 2.0 mHz ($4.9 \times 10^{-18}$). Combined with the systematic uncertainty of $5.7 \times 10^{-18}$ described in Table II, the total uncertainty for the clock comparison is $7.5 \times 10^{-18}$. The uptime rate for the comparison is mainly limited by the cryogenic optical clock.

## V. UP TIME RATE OF THE CLOCK

The uptime rate of the clock is measured, as shown in Fig. 4. The clock can reliably operate for over a week, with an uptime rate of over 95%. It is comparable to those optical clocks with long-term quasi-continuous performance, such as the Sr and Yb clocks mentioned above [30-32]. This experiment also verifies the reliability of the room-temperature optical clock after its preliminary transportation experiment from the optical table to the ground.

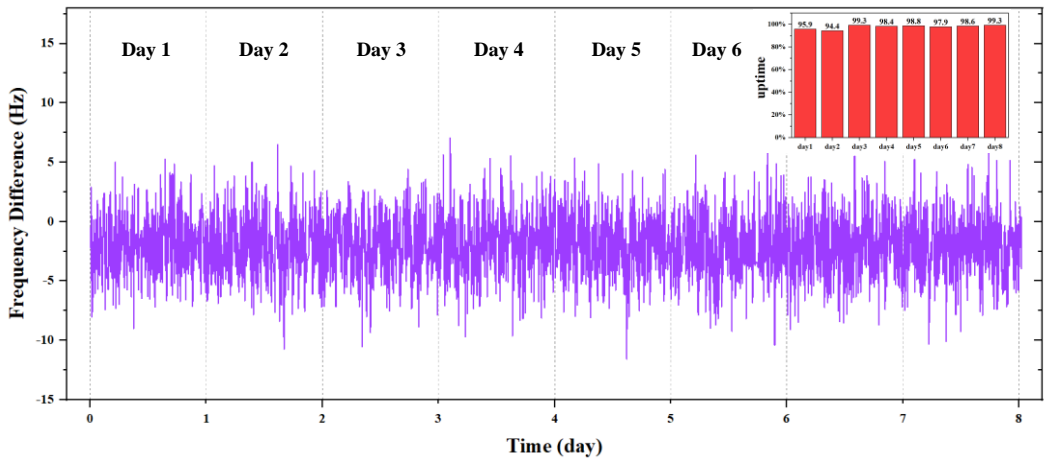

Fig. 4. The frequency difference between the transition pairs $^2S_{1/2}(m = \pm 1/2) \to {}^2D_{5/2}(m = \pm 1/2)$ and $^2S_{1/2}(m = \pm 1/2) \to {}^2D_{5/2}(m = \pm 3/2)$ of the room-temperature clock. Rabi interrogation method with a 20-ms probe time is used. The operating uptime rate reaches 97.8% in the period of around 8 days. The current continuous uptime rate of the room-temperature clock is limited by the ion loss caused by the instability of the frequency of the cooling laser and the interrogation operation of the probe laser, which occurs at a frequency of about once a day. Furthermore, the unlocking of the clock laser would also cause

the operation of the optical clock to be interrupted. Future improvements will focus on reloading the $^{40}$Ca$^+$ ion faster, and developing a more stable reference cavity for frequency stabilization of the clock laser.

## VI. CONCLUSION

In summary, we have developed a nearly continuous Ca$^+$ optical clock towards a transportable clock, with a systematic uncertainty of 4.8×10$^{-18}$ at room temperature. Through the optimization of the ion trap structure, the proportion of the effective solid angle of the insulating supports in the ion trap is greatly suppressed. Moreover, an active temperature control and evaluation scheme has been introduced. These efforts help to reduce the portable Ca$^+$ optical clock's BBR shift uncertainty to the 10$^{-18}$ level. The evaluation is also verified by measuring the frequency difference between the room-temperature and cryogenic clocks with an overall uncertainty of $7.5 \times 10^{-18}$. It is the first time that a transportable Ca$^+$ optical clock achieves the 10$^{-18}$ level systematic uncertainty and the first time that the Ca$^+$ optical clock comparison shows 10$^{-18}$ level total uncertainty. Higher accuracy clock comparisons can be implemented with optical clocks from other laboratories to test uncertainty evaluation and measurements of clock frequency ratios [28]. Our setups for reducing the BBR shift uncertainty are suitable for developing robust, transportable optical clocks based not only on Ca$^+$ ions, but also on other ions or neutral atoms.

## ACKNOWLEDGMENTS

Mengyan Zeng, Yao Huang contribute to this work equally. We thank Zhengtian Lu, Liyan Tang, Chaohui Ye, and Jun Luo for help and fruitful discussion. This work is supported by the National Key R&D Program of China (Grants No. 2022YFB3904001, 2022YFB3904004, 2018YFA0307500), the National Natural Science Foundation of China (Grants No. 12022414 and 12121004), the CAS Youth Innovation Promotion Association (Grants No. Y201963 and No. 2018364), the Natural Science Foundation of Hubei Province (Grant No. 2022CFA013) and CAS Project for Young Scientists in Basic Research (Grant No. YSBR-055).

\* guanhua@apm.ac.cn

† klgao@apm.ac.cn